# Spatial Antiferromagnetic Spin Texture as a Nano-Oscillator

**Victor S. Gerasimchuk[1], Yuri I. Gorobets[1,2], Oksana Yu. Gorobets[1,2], and Igor V. Gerasimchuk[1,2,*]**

[1]*Faculty of Physics and Mathematics, National Technical University of Ukraine "Igor Sikorsky Kyiv Polytechnic Institute", Peremohy Ave. 37, Kyiv 03056, Ukraine*

[2]*Department of Physics of Meso- and Nanocrystal Magnetic Structures, Institute of Magnetism, National Academy of Sciences of Ukraine and Ministry of Education and Science of Ukraine, Vernadsky Blvd. 36b, Kyiv 03142, Ukraine*

## ABSTRACT

We report a theoretical study of the localized spatial magnetization configuration, which is a confined spin configuration of the target skyrmion/hopfion type in an antiferromagnet with perpendicular magnetic anisotropy, and then we solve the particular problem of self-oscillations of such a topological spin texture. Using the energy approach, a self-consistent account of inhomogeneity of the characteristics of the topological magnetic spin texture was carried out. On this basis, the equation of free oscillations of the confined spin configuration magnetization was derived and its quasi-classical solution was found. For a thin ring spin texture, the frequency, period of oscillations and relative amplitude of the main tone of oscillations are found. For the first time, we determined the topological mass, inertial mass and total energy of the main tone of oscillations of such spatial spin texture. The self-oscillatory process of a spatial spin texture is interpreted as a magnetic nano-oscillator.

Two-dimensional solitons in magnets – skyrmions and vortices, topologically protected inhomogeneities of the magnetization field – are promising as information carriers for magnonics and spintronics [1–4]. Stable magnetic skyrmions were observed at room temperature and zero external magnetic fields in antiferromagnets (AFM) with perpendicular magnetic anisotropy [5,6]. It is AFM skyrmions that are considered the best carriers of information for data storage devices.

Let us pay attention to axisymmetric target skyrmions – spin configurations with a multi-ring structure, constant chirality, and continuous rotation of the $z$ component of the magnetization. Comparison of the results of numerical simulation with experimental measurements [7] demonstrates switching between two states of target skyrmions with opposite polarities and directions of rotation. Stable spatially localized states are also demonstrated by 3D topological solitons called hopfions. These axisymmetric variations of magnetization, closed and interlinked within the magnetic bulk of a toroidal shape, turns out to be energetically more favorable than chiral magnetic skyrmions in a wide range of parameters [8]. Numerical simulations demonstrated the transformation of the target skyrmion into a hopfion [9] and the existence of static solitons with different Hopf index in magnetic nanostructures with perpendicular magnetic anisotropy [10].

Possible spin configurations realized in magnets are not limited to skyrmions and hopfions. Recently, it was shown that in fact a standard model of chiral magnet possesses an infinite set of skyrmion solutions with different value and sign of topological charge and diverse morphology [11].

The dynamics of magnetic spin textures seems to be quite complex. Thus, skyrmions exhibit several well-known dynamic regimes, among them are breathing modes when the skyrmion radius changes [12,13], and rotational modes [14,15] when the position of the skyrmion core fluctuates. Various combinations of these modes and transitions between them [16], as well as inertial dynamics [17–19], can take place. Of course, more complex spin textures, i.e., those breaking more symmetries, have potentially more complex dynamics and eigenmodes.

To describe the oscillations of skyrmions and their translational motion, as a rule, the Thiele equation [20] is used, in which inertial effects are taken into account when introducing the inertial mass of a skyrmion [18,19]. However, the real oscillations of spin textures, which take into account the nonuniform distribution of the density of solitons' inertial masses and the densities of driving forces, have not been sufficiently studied. Oscillations of spin textures as a physical process are almost not discussed.

In this Letter, we intend to introduce a new research topic in this highly promising area. Due to the practical necessity of studying the dynamics of spatial topological spin configurations, we solve the problem of analytical calculation of eigenexcitations of an axisymmetric circular spin texture in a thin AFM layer with perpendicular magnetic anisotropy in the absence of applied external magnetic field.

Let us consider a two-sublattice AFM with the easy-axis magnetic anisotropy, assuming that $|\mathbf{M_1}| = |\mathbf{M_2}| = M_0 = \text{const}$, where $\mathbf{M_1}$ and $\mathbf{M_2}$ are the magnetizations of the AFM sublattices. The antiferromagnetism vector $\mathbf{L}$ and magnetization vector $\mathbf{M}$ of the magnetic system are defined via $\mathbf{M_1}$ and $\mathbf{M_2}$ as follows: $\mathbf{L} = \mathbf{M_1} - \mathbf{M_2}$ and $\mathbf{M} = \mathbf{M_1} + \mathbf{M_2}$. Consequently, we can write $\mathbf{M} \cdot \mathbf{L} = 0$, $\mathbf{M}^2 + \mathbf{L}^2 = 4 M_0^2$. The ground state of a two-sublattice AFM in the absence of an applied magnetic field corresponds to the complete mutual compensation of the magnetizations of the sublattices: $\mathbf{M_1} = -\mathbf{M_2}$, $\mathbf{M} = 0$ and $|\mathbf{L}| = L_0 = 2 M_0$.

The excited states of a two-sublattice AFM in the absence of dissipation are described by the Landau–Lifshitz equations in the following form [21,22]:

$$\begin{cases} \dfrac{\partial \mathbf{M}}{\partial t} = -g \cdot \left\{ \left[ \mathbf{M} \times \mathbf{H}_\mathbf{M}^{\text{eff}} \right] + \left[ \mathbf{L} \times \mathbf{H}_\mathbf{L}^{\text{eff}} \right] \right\}, & \mathbf{H}_\mathbf{M}^{\text{eff}} = -\dfrac{\delta F}{\delta \mathbf{M}}, \\ \dfrac{\partial \mathbf{L}}{\partial t} = -g \cdot \left\{ \left[ \mathbf{L} \times \mathbf{H}_\mathbf{M}^{\text{eff}} \right] + \left[ \mathbf{M} \times \mathbf{H}_\mathbf{L}^{\text{eff}} \right] \right\}, & \mathbf{H}_\mathbf{L}^{\text{eff}} = -\dfrac{\delta F}{\delta \mathbf{L}}, \end{cases} \quad (1)$$

where $g = 2\mu_0/\hbar$ is the gyromagnetic ratio, $\mu_0 = |e|\hbar/(2 m_e c)$ is the Bohr magneton, and $F$ is the AFM energy density functional.

In the case of a purely uniaxial AFM, the anisotropy axis of which coincides with the axis $Oz$, we obtain $L_x, L_y \ll L_z \approx |\mathbf{L}| = L_0 \approx 2 M_0$. In the natural for an AFM assumption $|\mathbf{M}| \ll |\mathbf{L}|$, magnetic energy density of an AFM can be written as follows:

$$F = \frac{A}{2}\mathbf{M}^2 + \frac{\alpha}{2} \sum_{i=1}^{3}\left(\frac{\partial \mathbf{L}}{\partial x_i}\right)^2 - \frac{\beta}{2} L_z^2 - (\mathbf{M} \cdot \mathbf{H}), \quad (2)$$

where $A$ is the homogeneous exchange energy constant, $\alpha$ is the inhomogeneous exchange energy constant, $\beta$ is the anisotropy constant, and $\mathbf{H}$ is the external magnetic field. Normally, $A \sim J/\mu_0 M_0$, where $J$ is the exchange integral, $\alpha \sim a^2 A$, $\alpha > 0$, where $a$ is the linear dimension of a crystal unit cell, and $\beta \ll A$. In Eq. (2) we omitted the terms with $\left(\partial \mathbf{M}/\partial x_i\right)^2$ in the inhomogeneous exchange energy and ignored the dependence of the anisotropy energy density on $\mathbf{M}$.

In the long-wave approximation, when the characteristic linear dimension of the inhomogeneity of the magnetization field is far beyond that of the linear dimension of a crystal unit cell, the second equation in (1) can be approximately rewritten as [22]

$$\frac{\partial \mathbf{L}}{\partial t} = g \cdot \left[ (\mathbf{H} - A \cdot \mathbf{M}) \times \mathbf{L} \right]. \qquad (3)$$

Equation (3) can be solved for $\mathbf{M}$:

$$\mathbf{M} = \frac{1}{4gAM_0^2} \cdot \left\{ \left[ \frac{\partial \mathbf{L}}{\partial t} \times \mathbf{L} \right] + g \cdot \left[ \mathbf{L} \times [\mathbf{H} \times \mathbf{L}] \right] \right\}, \qquad (4)$$

and the first equation in (1) can be rewritten as

$$\frac{\partial \mathbf{M}}{\partial t} = -g \cdot \left\{ \alpha \cdot [\mathbf{L} \times \Delta \mathbf{L}] + [\mathbf{M} \times \mathbf{H}] + \beta \cdot L_z \cdot [\mathbf{L} \times \mathbf{e_z}] \right\}, \qquad (5)$$

where $\mathbf{e_z}$ is the unit vector of the axis $Oz$. Substituting (4) into (5), we come to the equation for the AFM vector $\mathbf{L}$:

$$\left[ \mathbf{L} \times \left( c^2 \Delta \mathbf{L} - \frac{\partial^2 \mathbf{L}}{\partial t^2} \right) \right] = 2g \cdot (\mathbf{L} \cdot \mathbf{H}) \cdot \frac{\partial \mathbf{L}}{\partial t} + g^2 \cdot (\mathbf{L} \cdot \mathbf{H}) \cdot [\mathbf{L} \times \mathbf{H}] - \omega_0^2 L_z \cdot [\mathbf{L} \times \mathbf{e_z}]. \qquad (6)$$

Here $c = 2gM_0\sqrt{A \cdot \alpha}$ is the characteristic velocity which is equal to the minimum phase velocity of spin waves of the linear theory for $\mathbf{H} = 0$ (see, for ex., [21-24]), and $\omega_0 = c/\ell_0 = 2gM_0\sqrt{A \cdot |\beta|}$, where $\ell_0 = \sqrt{\alpha/|\beta|}$ is the characteristic magnetic length.

Equation (6) can be conveniently written in angular variables $\theta$ and $\varphi$ for the AFM vector $\mathbf{L} = L_0 \cdot \{\sin\theta \cdot \cos\varphi \cdot \mathbf{e_x} + \sin\theta \cdot \sin\varphi \cdot \mathbf{e_y} + \cos\theta \cdot \mathbf{e_z}\}$. Using vector $\mathbf{L}$ parameterization and assuming that the magnetic field $\mathbf{H}$ is directed along the anisotropy axis ($\mathbf{H} = H \cdot \mathbf{e_z}$), Eq. (6) can be written in the form of the system of dynamic equations for the angular variables:

$$\begin{cases} \dfrac{\partial^2 \theta}{\partial t^2} - c^2 \Delta \theta - \left[ \left( \dfrac{\partial \varphi}{\partial t} - \omega_H \right)^2 - c^2 (\nabla \varphi)^2 - \omega_0^2 \operatorname{sgn} \beta \right] \sin\theta \cos\theta = 0, \\ \dfrac{\partial}{\partial t} \left[ \sin^2 \theta \cdot \left( \dfrac{\partial \varphi}{\partial t} - \omega_H \right) \right] - c^2 \operatorname{div} \left( \sin^2 \theta \cdot \nabla \varphi \right) = 0, \end{cases} \qquad (7)$$

where $\omega_H = gH$.

The system of equations for a uniaxial two-sublattice AFM in the form analogous to (7) was obtained by Bar'yakhtar and Ivanov [23,24]. Some multidimensional vortex solutions in models of easy-axis and isotropic AFM were found in Ref. [25]. A new class of self-similar 3D nonlinear solutions of the system of equations (7) was derived in Ref. [26]. Three cases were investigated therein, viz., the spin wave velocity $\upsilon$ is less, greater and equal to the characteristic velocity $c$. We will use the following nonlinear solution of the system (7) [26]:

$$\begin{cases} \tan\dfrac{\theta}{2} = \sqrt{\dfrac{1-\mathrm{sn}(P(X,Y,Z),q)}{1+\mathrm{sn}(P(X,Y,Z),q)}}, \\ \varphi = \omega_H t + Q(X,Y,Z), \end{cases} \quad (8)$$

obtained in the case when the spin wave velocity is less than the characteristic velocity, $\upsilon < c$, and the Lorentz-like transformation of coordinates is applied: $X = x$, $Y = y$, $Z = z/\sqrt{1-\upsilon^2/c^2}$. This transformation means that Eqs. (7) are Lorentz invariant regarding coordinates and any nonlinear solution of these equations can move straightly with a constant velocity along the axis $Oz$.

The solution for the functions $P$ and $Q$ can be written in the form

$$\begin{cases} P = \chi \cdot (z-\upsilon t) + p(x,y), \\ Q = Q(x,y), \end{cases} \quad (9)$$

where parameter $\chi = \omega_0 / \left( c \cdot \sqrt{1-\upsilon^2/c^2} \right) = 1 / \left( \ell_0 \cdot \sqrt{1-\upsilon^2/c^2} \right)$, and the functions $p$ and $Q$ should satisfy the Cauchy–Riemann equations $\partial p/\partial x = -\partial Q/\partial y$, $\partial p/\partial y = \partial Q/\partial x$.

The relations (8), (9) define the indicated solution for $\theta$ and $\varphi$. Let us choose a certain form of the solution (8), (9) defining the function $p(x,y)$ as follows:

$$p = \ln\frac{r}{\tilde{r}_0}, \quad (10)$$

where $r = \sqrt{x^2+y^2}$ and $\tilde{r}_0$ is an arbitrary parameter of the dimension of length. Then from the Cauchy–Riemann equations we find the solution for the function $Q(x,y)$: $Q = -\alpha + \mathrm{const}$, where the parameter $\alpha = \arctan(y/x)$. As a result, solution (8) takes the final form:

$$\begin{cases} \tan\dfrac{\theta}{2} = \sqrt{\dfrac{1-\operatorname{sn}\left[\chi\cdot(z-\upsilon t)+\ln\left(r/\tilde{r}_0\right),q\right]}{1+\operatorname{sn}\left[\chi\cdot(z-\upsilon t)+\ln\left(r/\tilde{r}_0\right),q\right]}}, \\ \varphi = \omega_H t - \alpha + \text{const.} \end{cases} \quad (11)$$

In the limit $q \to 1$, that corresponds to the transition $\operatorname{sn}(u, q \to 1) \to \tanh(u)$, solution (11) is simplified:

$$\begin{cases} \theta = 2\cdot\arctan\dfrac{e^{-\chi(z-\upsilon t)}}{r/\tilde{r}_0}, \\ \varphi = \omega_H t - \alpha + \text{const.} \end{cases} \quad (12)$$

Let us discuss the main topological parameters of such a spin texture. The variation in spin configuration along the axis of symmetry, coinciding with the $z$ axis, is bounded on the $z$ axis. Indeed, an arbitrary constant $\tilde{r}_0$ in (10)–(12) affects only the shift of the spin texture as a whole along the $z$ axis. It follows from (12) that the size of the spin texture along the $z$ axis depends only on the characteristic magnetic length $\ell_0$. Therefore, by appropriately choosing the parameter $\ell_0$, one can adjust the thickness of the spin texture. Numerical analysis of solution (12) testifies that at the selected value of $\ell_0 = 0.7$ nm, the turn of the polar angle $\theta$ by $-\pi$ (from $\theta = \pi$ at $z \simeq z_1 = -2.5$ nm to $\theta = 0$ at $z \simeq z_2 = 2.5$ nm) occurs at the thickness of the spin texture $d \approx 5$ nm. The selected value of $\ell_0 = 0.7$ nm assumes a small exchange energy constant or a large uniaxial anisotropy constant, which is achievable in AFM [27,28]. In some cases it is convenient to put $\tilde{r}_0 = 50\ell_0$, it allows the spin texture to be symmetrically positioned about the plane $z = 0$.

At a fixed angle $\theta$, we obtain from Eq. (12) the following dependence for $z = z(r)$ at the initial time $t = 0$: $z = -\ell_0 \cdot \ln\left[\tan(\theta/2)\cdot r/\tilde{r}_0\right]$, which defines the locus of points having the same polar angle (for any value of the angle $\alpha$).

Taking into account the limitation of the magnetization distribution in height (along the $z$ axis), we can consider the spin texture to be rather thin if its characteristic radius $r$ significantly exceeds its thickness $d$, i.e., the condition $d/r \ll 1$ is met.

As follows from the above, in our case, a Néel-type spin configuration can be realized, in which the magnetization is oriented along the easy axis, and the rotation of atomic magnetic moments occurs in the film plane.

Let us define the surface density and the surface energy density of the magnet. Let us write the expression for the total energy of a uniaxial AFM [22]:

$$E = \int dV \left\{ \frac{\alpha M_0^2}{2c^2} \left[ c^2 (\nabla \theta)^2 + \left(\frac{\partial \theta}{\partial t}\right)^2 + \sin^2\theta \left( c^2 (\nabla \varphi)^2 + \left(\frac{\partial \varphi}{\partial t}\right)^2 - \omega_H^2 \right) \right] + \frac{\beta M_0^2}{2} \sin^2\theta \right\}, \quad (13)$$

where integration is taken over the volume of an AFM.

Using solution (12), from relation (13) in the approximation $\upsilon/c \ll 1$, we obtain the energy density of AFM in the form: $\varepsilon(r) = \varepsilon_0(r) + \varepsilon_k(r) \equiv \varepsilon_0(r) + \gamma(r) \cdot \upsilon^2/2$, where $\varepsilon_0(r)$ is the surface energy density, and $\varepsilon_k(r) = \gamma(r) \cdot \upsilon^2/2$ is the kinetic energy density of the spin texture. Hence, taking into account the ratio $\omega_0/c = 1/\ell_0 = \sqrt{|\beta|/\alpha}$, we find an expression for the surface density of the magnet:

$$\gamma(r) = \frac{r_0}{\sqrt{2}Ag^2} \left( \frac{1}{r_0^2} - \frac{1}{r^2} \right), \quad (14)$$

and surface energy density of the topological spin texture:

$$\varepsilon_0(r) = \frac{2\sqrt{2} \cdot \alpha M_0^2 r_0}{r^2}, \quad (15)$$

where parameter $r_0 = \ell_0/\sqrt{2}$.

Self-consistent account of inhomogeneity of the surface density and surface energy density is performed for the first time, and this makes it possible to estimate the value of the *spin texture mass* caused by its topology:

$$m_{\text{tp}} = \int_0^{2\pi} \int_{r_1}^{r_2} \gamma(r) r \, dr \, d\varphi = \frac{\pi m_0}{2} \left( \frac{r_2^2 - r_1^2}{r_0^2} - 2\ln\frac{r_2}{r_1} \right), \quad (16)$$

where $m_0 = \ell_0/(Ag^2) = 4\alpha M_0^2 \ell_0/c^2$. Indeed, the skyrmion spin texture can be described as a deformable quasiparticle [18], into the equation of motion of which the so-called "topological mass" is introduced, due to the topology of the structure.

From the positiveness of expression (14), it follows, in particular, the condition of applicability of this model with a function $p(x, y)$ in the form (10), namely $r > r_0$.

Let us use solution (12) for the analytical study of *axisymmetric oscillations of circular nanoscale spin texture* in AFM film with easy-axis magnetic anisotropy.

Let us approximate axisymmetric circular spin texture by a thin inhomogeneous membrane of variable density, the free oscillations of which are described by the wave equation

$$\gamma(r)\frac{\partial^2 u}{\partial t^2} = \operatorname{div}\left(\varepsilon_0(r)\cdot \operatorname{grad} u\right). \qquad (17)$$

In the polar coordinate system, in the simplest case of free axisymmetric oscillations, the right side of Eq. (17), taking into account (15), is reduced to $\varepsilon_0(r)\cdot\left(\dfrac{\partial^2 u}{\partial r^2} - \dfrac{1}{r}\cdot\dfrac{\partial u}{\partial r}\right)$. As a result, we obtain

*the equation of free oscillations of magnetization with inhomogeneous coefficients*:

$$\frac{\partial^2 u}{\partial t^2} = \frac{\varepsilon_0(r)}{\gamma(r)}\cdot\left(\frac{\partial^2 u}{\partial r^2} - \frac{1}{r}\cdot\frac{\partial u}{\partial r}\right), \qquad (18)$$

where the ratio $\varepsilon_0(r)/\gamma(r)$ characterizes the "magnetic rigidity" of the spin texture and is expressed in terms of the characteristic velocity $c$:

$$\frac{\varepsilon_0(r)}{\gamma(r)} = \frac{c^2}{(r/r_0)^2 - 1}.$$

We write Eq. (18) in the form of a wave equation with variable coefficients,

$$\frac{1}{c^2}\frac{\partial^2 u}{\partial t^2} = \frac{1}{(r/r_0)^2 - 1}\cdot\left(\frac{\partial^2 u}{\partial r^2} - \frac{1}{r}\cdot\frac{\partial u}{\partial r}\right), \qquad (19)$$

and formulate the problem of intrinsic excitations of an axisymmetric circular spin texture caused by initial perturbations. Let us consider a confined configuration of magnetization as an axisymmetric spin isosurface. By virtue of solution (12) and the restriction $r > r_0$ arising from expression (14), the spin isosurface can be represented as a circular ring, similar, for example, to hopfion [8–11] with its toroidal topology or target skyrmion [7] with an inner radius of circular edge twists exceeding $r_0$ ($r_0 \ll 1$). The implementation of a spin isosurface bounded by concentric rings, apparently, is also possible in confined geometry, for example, in nanotubes.

The standard procedure for separating the variables $u(r,t) = R(r)\cdot T(t)$ applied to Eq. (19) leads to a linear differential equation for the time component

$$\frac{d^2 T(t)}{dt^2} + (\lambda c)^2 T(t) = 0 \qquad (20)$$

and to the equation for the radial component

$$r\frac{d}{dr}\left(\frac{1}{r}\frac{dR(r)}{dr}\right)+(\lambda r)^2\left(\frac{1}{r_0^2}-\frac{1}{r^2}\right)R(r)=0 \tag{21}$$

with the separation constant $\lambda$.

Equation (20) has obvious oscillating solutions, but the exact solution of Eq. (21) is not known in general. However, for a new unknown function $y(x) = R(r)$ with a dimensionless variable $x = \lambda r^2/r_0$, Eq. (21) can be reduced to an equation of the type of the stationary Schrödinger equation,

$$\frac{d^2 y}{dx^2}+\left(\frac{1}{4}-\frac{\lambda r_0}{4x}\right)y=0, \tag{22}$$

with the potential $U(x) = \lambda r_0/(4x)$.

The quasi-classical consideration, in particular, the WKBJ method is applicable to Eq. (22). Following [29,30], we assume that the changes of $U(x)$ at the wavelength are so small that over several wavelengths the potential $U(x)$ can be considered to be constant in the vicinity of the turning point $x = x_0$. For the effective wavenumber $q(x) = (1/4 - U(x))^{1/2}$, the condition for a slow change of the potential $U(x) = \lambda r_0/(4x)$ at a wavelength (*the quasi-classical condition*) at small values $|x - x_0|$ is determined by the inequality

$$|x - x_0| \gg x_0^{1/3}. \tag{23}$$

One more condition for $q(x)$, namely $q(x) = 0$ [29] defines the isolated classical turning point $x = x_0$:

$$1 - \frac{\lambda r_0}{x} = 0, \qquad x_0 = \lambda r_0 \qquad (x_0 > 0). \tag{24}$$

Thus, Eq. (22) (and along with it Eq. (21)), in accordance with the WKBJ method [29,30], has a certain set of solutions: (i) in the immediate vicinity of the turning point (24), and (ii) some set of solutions to the right and left of the turning point, in the quasi-classical region defined by inequality (23). In the first case, the solution is written in terms of the Airy functions, and in the second one, the WKBJ method for the potential $U(x) \sim x^{-1}$ gives oscillating solutions to the right of the turning point and exponential solutions to the left of it. Note that Eq. (23) is consistent with the condition of applicability of the model under consideration, $r > r_0$.

Let us restrict ourselves here to the second case, which is more interesting from a practical point of view. Let the ring nanoscale spin isosurface bounded by two concentric circles lie entirely in the region of the quasi-classical solution of Eq. (22). We assume that the inner $x_1$ and outer $x_2$ radii of the ring spin configuration significantly exceed the distance to the turning point $x_0$ ($x_1, x_2 \gg x_0$). Then it follows from (23) that the ring spin isosurface lies entirely in the outer region (with respect to the turning point) of the quasi-classical approximation. The other quasi-classical region, internal with respect to the turning point, is a circle of radius $x_0$ centered at the origin of coordinates. However, we are not interested in this region here.

The magnetization configuration (12) at the initial time $t = 0$ for the magnetic length $\ell_0 = 0.7$ nm, parameter $\tilde{r}_0 = 50\ell_0$ and the thickness of the spin texture $d \approx 5$ nm is shown in Figs. 1(a) and 1(b). Figure 1(a) presents the ring spin texture, and Figure 1(b) demonstrates the turn of the polar angle $\theta$ by $-\pi$ (from $\theta = \pi$ at $z \simeq -2.5$ nm to $\theta = 0$ at $z \simeq 2.5$ nm) at a fixed radius between inner and outer ones.

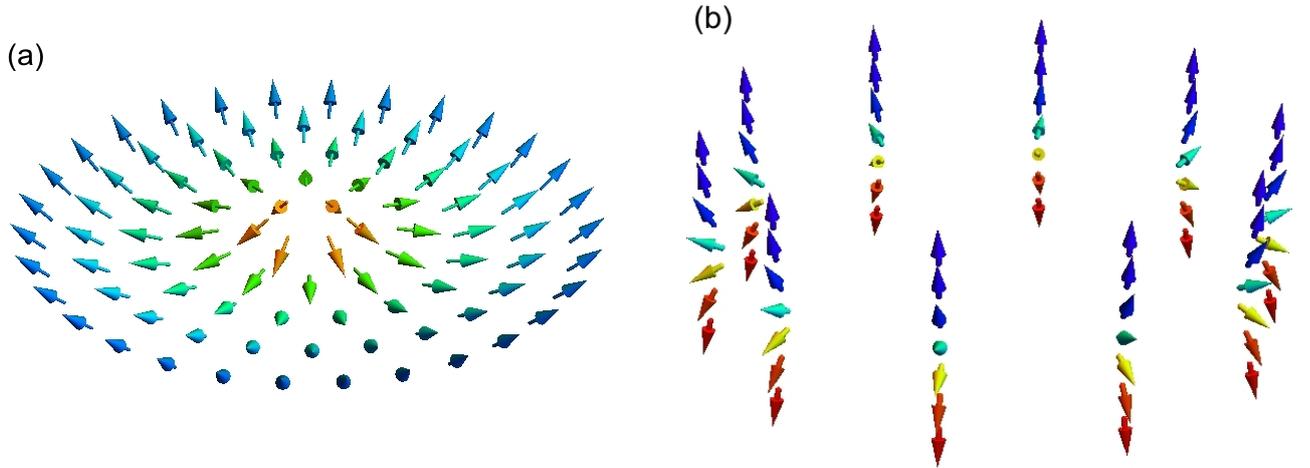

FIG. 1. Ring spin configuration (12) at the initial time $t = 0$ for the magnetic length $\ell_0 = 0.7$ nm, parameter $\tilde{r}_0 = 50\ell_0$ and the thickness of the spin texture $d \approx 5$ nm. (a) Ring spin texture. (b) Magnetization configuration and turn of the polar angle $\theta$ by $-\pi$ from $\theta = \pi$ at $z \simeq -2.5$ nm to $\theta = 0$ at $z \simeq 2.5$ nm for a fixed radius between inner and outer ones.

The boundary value problem in the region under consideration ($x \gg x_0$) is formulated, for example, as follows: to find nontrivial solutions of Eq. (22) that satisfy the condition of fixing spin isosurface on circles of inner and outer radii,

$$y(x_1) = 0, \qquad y(x_2) = 0 \qquad (x_1 < x < x_2), \qquad (25)$$

caused by the radially symmetric initial perturbation

$$u(x,0) = -4h \frac{x - \sqrt{x}\left(\sqrt{x_1} + \sqrt{x_2}\right) + \sqrt{x_1 x_2}}{\left(\sqrt{x_2} - \sqrt{x_1}\right)^2}, \qquad u_t(x,0) = 0, \qquad (26)$$

where $x_i = \lambda r_i^2 / r_0$ $(i = 1, 2)$. Conditions (26) correspond to the initial profile of the ring spin isosurface in the form of a paraboloid of revolution with height $h$ $(h \ll x_1, x_2)$, which is symmetric with respect to the middle of the ring at zero initial oscillation velocities.

If we expand the potential $U(x)$ in Eq. (22) near $x = x_0$ in a power series of $|x - x_0|$, restricting ourselves to the linear term of the expansion, then the square of the effective wavenumber $q^2(x) \simeq (x - x_0)/(4\lambda r_0)$ and Eq. (22) is reduced to the differential equation

$$\frac{d^2 y}{dx^2} + \frac{1}{4\lambda r_0}(x - x_0) y = 0 \qquad (27)$$

which has an asymptotic solution, determined by the formula [29]

$$y(x) \approx \sqrt{\frac{2}{\pi q}} \cdot \left[ A \cos\left(w - \frac{5}{12}\pi\right) + B \cos\left(w - \frac{1}{12}\pi\right) \right], \qquad (28)$$

where $w(x) = \int_{x_0}^{x} q(x) dx$.

Having required the fulfillment of boundary conditions (25) for the solution (28), we derive a homogeneous system of linear equations, from the condition of non-trivial compatibility of which we obtain the equation

$$\sin(w_1 - w_1) = \sin \frac{\lambda(\rho_1 - \rho_2)}{3 r_0^2} = 0$$

for finding the eigenvalues $\lambda$:

$$\lambda_n = \pi n \frac{3 r_0^2}{\rho_1 - \rho_2}, \qquad n = 1, 2, \ldots, \qquad (29)$$

where $\rho_i \equiv \left(r_i^2 - r_0^2\right)^{3/2}$ $(i = 1, 2)$.

Using solution (28), we find the eigenfunctions corresponding to the eigenvalues (29):

$$y\left(x = \frac{\lambda}{r_0}r^2\right) \approx A\sin\frac{\lambda_n(\rho-\rho_2)}{3r_0^2}. \tag{30}$$

As a result, we obtain a solution of the eigenvalue problem for Eq. (21) and boundary conditions (25):

$$R_n(r) \approx \sqrt{\frac{3r_0}{\pi\rho^{1/3}}}\frac{A}{\cos\left(\frac{\lambda_n\rho_2}{3r_0^2}-\frac{\pi}{12}\right)}\sin\frac{\lambda_n(\rho-\rho_2)}{3r_0^2}, \tag{31}$$

where we used $q(r) = \rho^{1/3}/(2r_0)$.

Taking into account the eigenvalues (30) of the boundary value problem, we write the solution of the linear differential equation (20):

$$T_n(t) = A_n\cos(\lambda_n ct) + B_n\sin(\lambda_n ct),$$

and the general solution of the wave equation (19):

$$u(r,t) = \sqrt{\frac{3r_0}{\pi\rho^{1/3}}}\cdot\sum_{n=1}^{\infty}\left[A_n\cos(\lambda_n ct) + B_n\sin(\lambda_n ct)\right]\cdot\frac{\sin\frac{\lambda_n(\rho-\rho_2)}{3r_0^2}}{\cos\left(\frac{\lambda_n\rho_2}{3r_0^2}-\frac{\pi}{12}\right)}. \tag{32}$$

Now we require the fulfillment of the initial conditions (26). Based on the conditions of orthogonality and normalization of the main system of trigonometric functions, we obtain an expression for determining the coefficients $A_n$ (coefficients $B_n = 0$):

$$A_n = \widetilde{C}_n(r_0)\cdot U_n(\rho_1-\rho_2), \tag{33}$$

where

$$\widetilde{C}_n(r_0) = \sqrt{\frac{\pi}{3r_0}}\frac{1}{(\rho_1-\rho_2)}\cdot\cos\left(\frac{\lambda_n\rho_2}{3r_0^2}-\frac{\pi}{12}\right), \qquad \rho \equiv \left(r^2-r_0^2\right)^{3/2},$$

$$U_n(\rho_1-\rho_2) = \int_0^{2(\rho_1-\rho_2)} u(\rho,0)\cdot\rho^{1/6}\sin\frac{\lambda_n(\rho-\rho_2)}{3r_0^2}d\rho,$$

$$u(\rho,0) = -\frac{4h}{(r_1 - r_2)^2} \left[ \left(\rho^{2/3} + r_0^2\right) - (r_1 + r_2)\left(\rho^{2/3} + r_0^2\right)^{1/2} + r_1 r_2 \right].$$

The final solution of the boundary value problem for the ring spin isosurface (19), (25), (26), taking into account (33), takes the form

$$u(r,t) = \frac{1}{(\rho_1 - \rho_2)\rho^{1/6}} \cdot \sum_{n=1}^{\infty} U_n(\rho_1 - \rho_2) \cdot \cos(\lambda_n ct) \cdot \sin\frac{\lambda_n(\rho - \rho_2)}{3r_0^2}. \quad (34)$$

Coefficients $A_n$ (33) are not possible to calculate in an analytical form. With the selected characteristic magnetic length $\ell_0 = 0.7$ nm and the sizes of the ring spin texture $r_1 = 25$ nm and $r_2 = 50$ nm ($d/(r_2 - r_1) \ll 1$), the first 10 coefficients $U_n(\rho_1 - \rho_2)$ of series (34), normalized to $h$ (we denoted $U_{nh} \equiv U_n/h$, where $n = 1, 2, ..., 10$), lie in the interval

$$U_{1h} \simeq 3.130 \cdot 10^{-26} \, m^{7/2}, \ldots, U_{10h} \simeq 7.844 \cdot 10^{-28} \, m^{7/2}$$

and decrease with increasing $n$. Therefore, the solution of the boundary value problem, normalized to $h$, $u_h(r,t) \equiv u(r,t)/h$ can be presented in the form

$$u_h(r,t) \approx \frac{1}{(\rho_1 - \rho_2)\rho^{1/6}} \cdot \sum_{n=1}^{10} U_{nh}(\rho_1 - \rho_2) \cdot \cos(\lambda_n ct) \cdot \sin\frac{\lambda_n(\rho - \rho_2)}{3r_0^2} \equiv \sum_{n=1}^{10} u_{nh}(r,t). \quad (35)$$

The dependence $u_h(r,t)$ (35) on the radius $r$ at the initial time $t = 0$ is presented in Fig. 2 (we put the characteristic velocity to be equal to $c = 5 \cdot 10^3$ m/s).

For the main tone of oscillations (the first term in (35) for $n = 1$), the obtained solution (35) allows finding the relative amplitude, frequency and period of oscillations:

$$u_{1h}(r,t) \simeq \frac{2.862 \cdot 10^{-4}}{\left(r^2 - r_0^2\right)^{1/4}} \cdot \cos(\lambda_1 ct) \cdot \sin\frac{\lambda_1(\rho - \rho_2)}{3r_0^2}, \qquad \lambda_1 \simeq 21113.276,$$

(36)

$$\omega_1 = \lambda_1 c \simeq 1.056 \cdot 10^8 \, s^{-1}, \qquad T_1 = \frac{2\pi}{\omega_1} \simeq 5.952 \cdot 10^{-8} \, s.$$

The oscillation frequency $\omega_1$ is of the order of magnitude corresponding to the frequencies of the oscillations of magnetic domain walls in confined geometries [31–33].

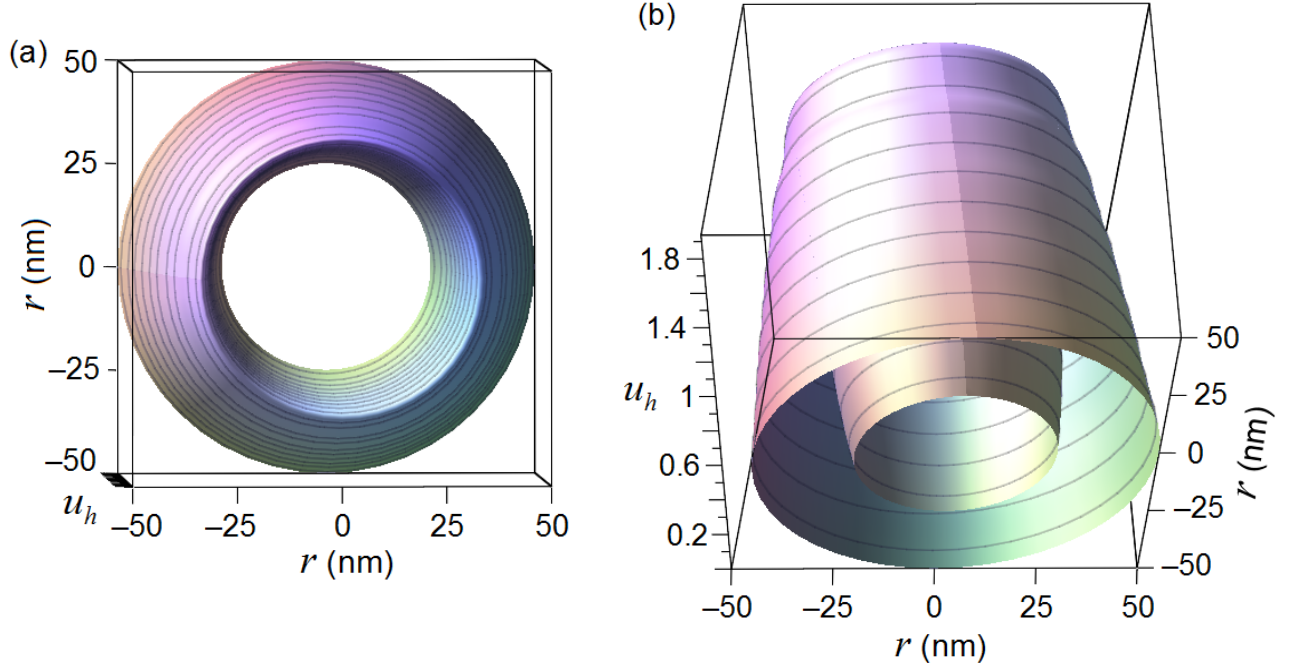

FIG. 2. Ring spin isosurface (35) at the initial time $t = 0$.

Using solution (34), one can estimate the total oscillation energy of the ring spin isosurface:

$$E = \frac{1}{2}\int_{r_1}^{r_2}\left[\gamma(r)\left(\frac{\partial u}{\partial t}\right)^2 + \varepsilon_0(r)\left(\frac{\partial u}{\partial r}\right)^2\right]rdr \simeq 11.36 m_0 \sum_{n=1}^{\infty}\omega_n^2\frac{U_n^2(\rho_1-\rho_2)}{\ell_0(\rho_1-\rho_2)^2},$$

where the coefficient before the sum is identified with its mass

$$m_{\text{inert}} = 11.38\cdot m_0 = 11.38\cdot 4\alpha M_0^2\ell_0/c^2 \simeq 1.3\cdot 10^{-27}\text{ kg}. \qquad (37)$$

The resulting value gives an estimate of the mass due to the axisymmetric oscillations of the spin texture and which can be interpreted as the *inertial mass* of the spin isosurface. The value of inertial mass (37) correlates with the corresponding value at the microscopic description of the kinematic properties of skyrmions [34] and antiferromagnetic domain walls in antiferromagnet–heavy-metal bilayers [35].

In the general case, as noted in [18], the inertial mass differs from the topological mass. In our case, the inertial mass $m_{\text{inert}}$ indeed has an order of magnitude different from the *topological mass* $m_{\text{tp}}$ calculated by the formula (16):

$$m_{\text{tp}} = \frac{\pi m_0}{2}\left(\frac{r_2^2 - r_1^2}{r_0^2} - 2\ln\frac{r_2}{r_1}\right) \simeq 6.6\cdot 10^{-24}\text{ kg}. \qquad (38)$$

The value of the topological mass (38) is consistent with a typical mass scale of skyrmion spin texture [17,28] and with the mass of geometrically confined magnetic domain walls [31,36,37].

In conclusion, we proposed a concept of the spatial distribution of magnetization, which is a confined spin configuration of the target skyrmion/hopfion type in an antiferromagnet with perpendicular magnetic anisotropy. Using a three-dimensional inhomogeneous antiferromagnetic magnetization distribution and an energy approach, the surface energy density of the topological spin texture and surface density of the magnet are calculated as functions of the radius of spin isosurface. Taking these characteristics into account, we derived the equation of free oscillations of topological spin texture magnetization and found its quasi-classical solution. For a thin axisymmetric circular nanoscale spin configuration, the main characteristics of eigenexcitations are determined, viz., frequency, relative amplitude and period of oscillations. Moreover, we found the topological and inertial masses of such a localized spin texture which correlate with the known numerical and experimental results.

The localized oscillating spin texture is not only a source of rich physics, but also a promising candidate for advances in spintronics. We treat the self-oscillatory process of a localized spin texture as a *magnetic nano-oscillator* and recommend a tunable radio-frequency (HF) generator as a new mechanism.

I.V.G. acknowledges support by the Ministry of Education and Science of Ukraine (Project No. 0121U110090).

*Corresponding author.
igor.gera@gmail.com